\documentclass[useAMS,usenatbib]{mn2e}
\usepackage{psfig}
\usepackage{amssymb}
\usepackage{color}

\title[Fiber Assignment in Next-generation Wide-field Spectrographs]
{Fiber Assignment in Next-generation Wide-field Spectrographs}

\author[I. Morales, A. D. Montero-Dorta, M. Azzaro, F. Prada, J. S\'anchez \& S. Becerril]{
\parbox[t]{\textwidth}{
Isaac Morales$^1$, Antonio D. Montero-Dorta$^{1,}$\thanks{E-mail: amdorta@iaa.es}, Marco Azzaro$^1$, Francisco Prada$^{1,2}$, Justo S\'anchez$^1$ \& Santiago Becerril$^1$}
\vspace*{6pt} \\ 
$^1$Instituto de Astrof\'isica de Andaluc\'ia (CSIC), Granada, E-18008, Spain \\
$^{2}$Visiting Professor at the  Institute for Computational Cosmology, Department of Physics, University of Durham, UK 
\vspace{-0.5cm} 
}

\date{Accepted ---. Received ---;in original form ---}

\newcommand{\plotone}[1]
           {\centering \leavevmode \psfig{file=#1,width=\columnwidth,clip=}}

\def\simlt{\lower.5ex\hbox{$\; \buildrel < \over \sim \;$}}
\def\simgt{\lower.5ex\hbox{$\; \buildrel > \over \sim \;$}}
\usepackage{graphicx}
\usepackage{rotating}

\definecolor{red}{rgb}{1,0,0}

\begin{document}

\bibliographystyle{mnras}

\maketitle


\begin{abstract}

We present an optimized algorithm for assigning fibers to targets in 
next-generation fiber-fed multi-object spectrographs. The method, that we named
draining algorithm, ensures that the maximum number of targets in
a given target field is observed in the first few tiles. Using randomly 
distributed targets and mock galaxy catalogs we have estimated that the 
gain provided by the draining algorithm as compared to a random assignment
can be as much as $2\%$ for the first tiles. This would imply 
for a survey like BigBOSS saving for observation several hundred 
thousand objects or, alternatively, reducing the covered area in $\sim 350~deg^{2}$. 
An important advantage of this method is that the fiber collision problem can be solved easily and in an optimal way.
We also discuss additional optimizations of the fiber positioning process. In particular,
we show that allowing for rotation of the focal plane can improve the efficiency 
of the process in $\sim 3.5 - 4.5 \%$ even if only small adjustments are permitted (up 
to $2^{o}$). For instruments that allow large rotations of the focal plane the 
expected gain increases to $\sim 5 - 6 \%$. These results, therefore, strongly support 
focal plane rotation in future spectrographs, as far as the efficiency of the
fiber positioning process is concerned. Finally, we discuss on the implications of 
our optimizations and provide some basic hints for an optimal survey strategy 
based on the number of targets per positioner.

\end{abstract}


\section{Introduction}
\label{sec:intro}

In the next decade astronomers will be challenged to constrain the nature of  
dark matter, dark energy and the perhaps inflationary processes  
which generated structure in the Universe. Much effort will also be devoted 
to shedding light into the astrophysics of galaxy formation and evolution. Observationally, 
large-scale galaxy surveys have consolidated, since the pioneering Center for Astrophysics Redshift Survey 
(CfA RS, \citealt{Huchra1983}), as the most efficient way to approach these and other fundamental 
studies. Currently, the complexity and the scale of the phenomena involved in these studies demand
progressively larger and deeper galaxy samples. In addition, more accurate measurements of 
fundamental physical properties from astronomical objects (such as kinematics, 
temperature, gravity/mass, chemical abundances or age) are needed. 
These requirements can only 
be satisfied through spectroscopy, which has become critical to further astrophysical understanding. 
There is a growing awareness in the community that answering many of the
pressing astrophysical and cosmological questions of the coming decades 
will require undertaking vast deep spectroscopic surveys, mainly on  
large aperture telescopes (see \citealt{Peacock2006, Bell2009}). 
The development and optimization of spectroscopic survey techniques and strategies
represents therefore a necessary step in order to increase the efficiency of our experiments and
provide theorists with more accurate observational constraints.\\ 

Multi-object spectroscopy can be performed with traditional slits or modern optical
fibers. Both approaches present advantages and disadvantages, but the latter provides much more 
versatility in terms of object collection and spectral coverage, which makes it more suitable for large-scale 
spectroscopic facilities. A number of fiber-fed spectrographs intended for large-scale galaxy surveys 
have been proposed recently. A good example is the Wide-Field Multi-Object Spectrograph (WFMOS, \citealt{Bassett2005}), which was proposed 
for the 8.2-m Subaru Telescope and aimed for a detailed investigation into the nature 
of dark energy and into galaxy formation and evolution. Similar scientific
motivations supported the proposal of a wide-field fiber-fed spectrograph, the 
Super Ifu Deployable Experiment (SIDE, \citealt{Prada2008}), for the 10-m Gran Telescopio Canarias.
None of these instruments were finally built but they are excellent examples of 
state-of-the-art survey spectrographs aiming to fulfill next-generation scientific requirements.
Several spectrographs have otherwise been accepted for construction, such as the Fibre Multi-Object 
Spectrograph (FMOS, \citealt{Kimura2010}), a near-infrared instrument which is already 
mounted on the Subaru Telescope. Already in the last stages of commissioning is also the set of 16 multi-fiber spectrographs for the 
five-degree field of view Chinese Large Sky Area Multi-Object Fibre Spectroscopic Telescope (LAMOST, \citealt{Wang2009}). LAMOST
was conceived to carry out several wide-field spectroscopic surveys focusing on both the 
structure of the Milky Way and the large-scale structure of the Universe. Finally, as an example 
of next-generation spectroscopic facilities, we will mention the Big 
Baryonic Oscillation Spectroscopic Survey (BigBOSS, \citealt{Schlegel2009}). BigBOSS is a proposed 
ground-base dark energy experiment to study baryon acoustic oscillation with an all-sky galaxy redshift survey, making
use of a multi-object fiber-fed spectrograph on the Mayall 4-m telescope at KPNO.\\

For all fiber-fed spectrographs (as those mentioned above) a primordial and common technical
problem is the positioning of fiber ends, which must match the projected position of objects in 
the focal plane. In order to solve this difficulty, the concept used in most 
recently-proposed fiber-fed spectrographs (SIDE, LAMOST, BigBOSS, etcetera) consists 
of an array of fiber positioners covering the entire focal plane which is able to position all fiber heads 
simultaneously. This solution reduces drastically the reconfiguration time of the system as compared to the 
most common alternative based on a pick-and-place device. Important for this work, in a real survey normally 
a single configuration of the array of positioners is not enough to observe all (or even a given required fraction) of the objects
in a target field. Several configurations (or tiles) are needed to reach a given completeness, depending on
the typical number of objects per fiber positioner. The purpose of this work is to present an optimized
way to assign fibers to objects so that the maximum number of objects is assigned in the first 
tiles. We also discuss on some additional ways of optimizing the 
fiber positioning process, depending on the capabilities of the instrument.\\ 

This paper is organized as follows. In Section~\ref{sec:nom}, we define some important concepts and provide the nomenclature that 
we use throughout this work. In Section~\ref{sec:robot}, we briefly describe
a general system consisting of an array of positioners covering the intrument focal plane. In Section~\ref{sec:draining}, 
we present our optimized fiber positioning algorithm, that we call draining algorithm, and 
assess its performance as compared to a random assignment in catalogs of randomly distributed
objects. Section~\ref{sec:additional} is dedicated to evaluating further optimizations 
such as rotation of the focal plane. In Section~\ref{sec:real}, we test the
efficiency of our optimizations in mock galaxy catalogs drawn from cosmological simulations. Finally, in Section~\ref{sec:discussion}, we            
summarize our main results and discuss on their implications.

\section{Nomenclature and definitions}
\label{sec:nom}

In order to facilitate the reader's comprehension we will first introduce
some basic concepts that will be frequently quoted in this work. Namely:

\begin{itemize}

\item {\bf{Target:}} Each of the objects that we plan to observe.

\item {\bf{Target field:}} An area on the sky that we plan to observe.    

\item {\bf{Target density:}} Total number of targets per sq. deg. in 
a given target field. 

\item {\bf{Position angle (PA):}} The angle that describes the rotation 
of the focal plane around an arbitrary fixed axis perpendicular to it.

\item {\bf{Tile:}} A spectrograph exposure at a given telescope pointing and
position angle.

\item {\bf{Tile density:}} Number of tiles per sq. deg. in a given target field. 

\item {\bf{Completeness:}} The minimum fraction of targets that we need to
observe in a given target field in order to satisfy the scientific requirements of the survey. 

\item {\bf{Tiling:}} The complete process of minimizing the number of tiles needed to 
observe a given target field with the requested completeness. This process not only 
involves finding the optimal number of tiles but also the position of each tile 
in the target field (see \citealt{Blanton2003}).  

\item {\bf{Positioner:}} A mechanical device for positioning a fiber in the desired 
location within a certain region of the focal plane of the instrument. The region 
that each positioner can patrol is called Patrol Disc.      

\item {\bf{Fiber density:}} Number of fibers (hence positioners) per sq. deg. in 
a given target field.

\item {\bf{Target-to-positioner ratio ($\eta$):}} The number of targets per positioner or, equivalently,
the target density divided by the fiber density. 

\item {\bf{Fiber collision:}} A conflict that occurs when two or more targets  
are so closely located in the focal plane that they cannot be observed simultaneously, due to the 
physical size of the positioners. The minimum separation that prevents a situation like this 
from happening depends on the geometry of the positioner.

\end{itemize}

Note that the main intention of this work is to present a complete and optimized method for fiber 
assignment. We will therefore, and unless otherwise stated, consider a target 
field of approximately the size of the focal plane of the spectrograph. Only 
rotation of the focal plane will be allowed as an additional optimization.  
The process of tiling itself, that involves positioning tiles on a large region 
of the sky (as compared to the size of a tile), will only be discussed in a qualitative way.

\section{Focal plane description}
\label{sec:robot}

\begin{figure}
\plotone{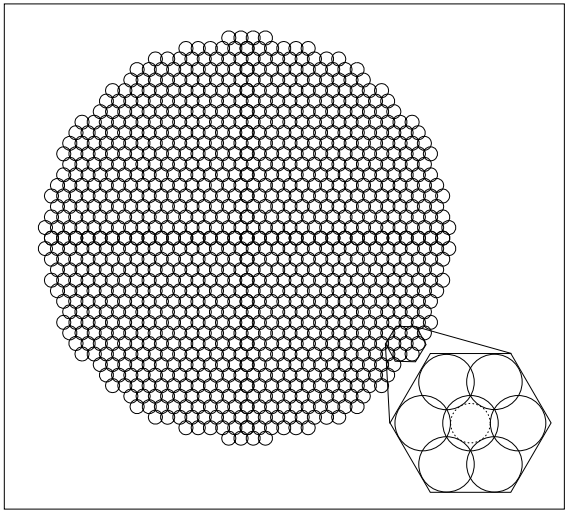}
\caption{Sketch of a fiber positioning robot represented by an array 
of patrol discs covering the entire focal plane. Each positioner of the robot 
is dedicated to one of these patrol discs. This example features 1003 patrol discs for 
the 992-mm diameter field of view at the GTC. The detail shows a group of 7 patrol discs which overlap
in a hexagonal pattern.}
\label{fig:focalplane}
\end{figure}

\begin{figure}
   \begin{center}
   \includegraphics[width=12cm]{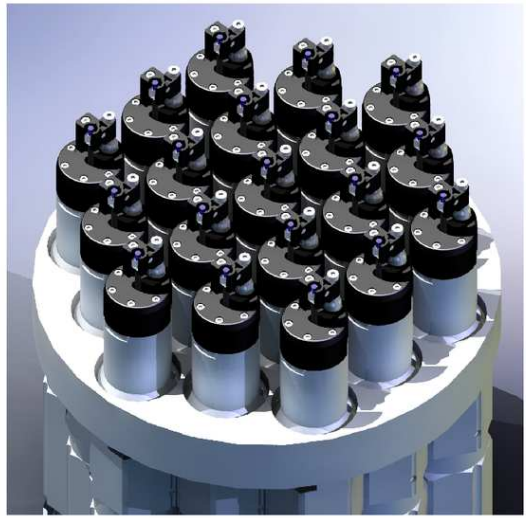}
   \end{center}
      \caption{A view of a subset of a positioner array mounted on its
holder. Here, 19 units are shown as an example.}
         \label{fig:19_positioners}
   \end{figure}

The concept that we outline here follows a standard design that can be 
extrapolated to most future fiber-fed multi-object spectrographs. In these 
instruments, the focal plane is populated with an array of fiber positioners, 
distributed in a hexagonal pattern, so that a single device is used to position each fiber head
in the desired location within its patrol disc. 
Each positioner is therefore devoted to observing a single target.
In a few words, the fiber positioning robot is a collection of
positioners, all identical, distributed over an array which covers the entire
focal plane. The focal plane is therefore covered
by these patrol discs so that all possible positions can be reached by at least
one positioner (see Fig.~\ref{fig:focalplane}). In order to cover the whole focal plane, 
a certain degree of overlap between patrol discs is needed so some regions 
of the focal plane can be reached by more than one positioner (two or maximum three), 
hence rising the possibility of fiber collisions. 
In Fig.~\ref{fig:19_positioners} we show a view of a real subset of a fiber positioner robot, 
with $19$ positioners in hexagonal pattern \citep{Azzaro2010}.\\

This concept offers a number of advantages as compared to others. It is 
robust, scalable and easy to service and maintain (failure of one 
positioner causes the loss of one target only). Operationally, it provides
extremely short reconfiguration times and allows an efficient real-time correction
of differential atmospheric dispersion. It is, however, specifically
conceived for wide-field surveys. The reason is that positioners cannot be
densely packed onto a small portion of the focal plane. Consequently, the system is efficient for
rather uniform distributions of targets or, alternatively, for large areas where any 
possible bias is smoothed. Find a complete discussion on this fiber positioning 
concept in \cite{Azzaro2010}. \\

The results presented in this work on the optimization of the fiber assignment process
are based on a complete simulation of the focal plane of the SIDE spectrograph, which adopted
a fiber positioning robot as that described above. SIDE is an example of a state-of-the-art fiber-fed instrument 
capable of efficiently undergoing next-generation large-scale spectroscopic surveys. The 
focal plane of the SIDE spectrograph was designed as an array of 1003 positioners covering the 
20-arcmin field of view at GTC. Important for this work, the results obtained with this simulation are 
valid for any focal plane consisting in an array of positioners as that outlined in this section. The key 
parameter to describe the efficiency of the fiber assignment process, as we will see below, is the 
target-to-positioner ratio, $\eta$ (instead of other parameters such as the size of the focal plane or the number of positioners). 
In order to better illustrate our results we will also discuss another real-life example: BigBOSS. In 
Table~\ref{tab:BB_SIDE} we list some of the relevant parameters for both SIDE and BigBOSS.\\

   \begin{table*}
      \caption{SIDE and BigBOSS focal plane parameters.}
         \label{tab:BB_SIDE}
         \centering
         \begin{tabular}{c | c c c c c}
            \hline
            \noalign{\smallskip}
                   &  Field of View    & Target density  & Number of positioners      &  Fiber density   & Target-to-positioner ratio ($\eta$)  \\
                   &  ($deg^{2}$)   & ($targets/ deg^{2}$)  &       &  ($fibers/ deg^{2}$)   &   \\
            \noalign{\smallskip}
            \hline
            \noalign{\smallskip}
            {\bf{SIDE}}   & 0.085   & 11500        & 1003  & 11500  & 1  \\
  
            \noalign{\smallskip}
            \hline
            \noalign{\smallskip}
            \noalign{\smallskip}
            {\bf{BigBOSS}}      & 7    & 3500  & 5000   & 714     & 5      \\
 
            \noalign{\smallskip}
            \hline
         \end{tabular}
   \end{table*}

\section{Fiber assignment algorithm}
\label{sec:draining}

In this section we briefly describe the philosophy behind our optimized fiber positioning algorithm: the
draining algorithm. We also discuss its performance as compared to a simple random approach. Let us
consider a generic instrument with a fiber positioning robot like
that described in the previous section covering 
the entire focal plane with $N$ positioners. A set of targets is used to 
populate the focal plane according to 
a given target-to-positioner ratio, $\eta$. We 
will first assume that targets are randomly distributed
in the focal plane. The design of the system considered
implies that each target is reachable by one, two or maximum three positioners.\\

\begin{figure}
\plotone{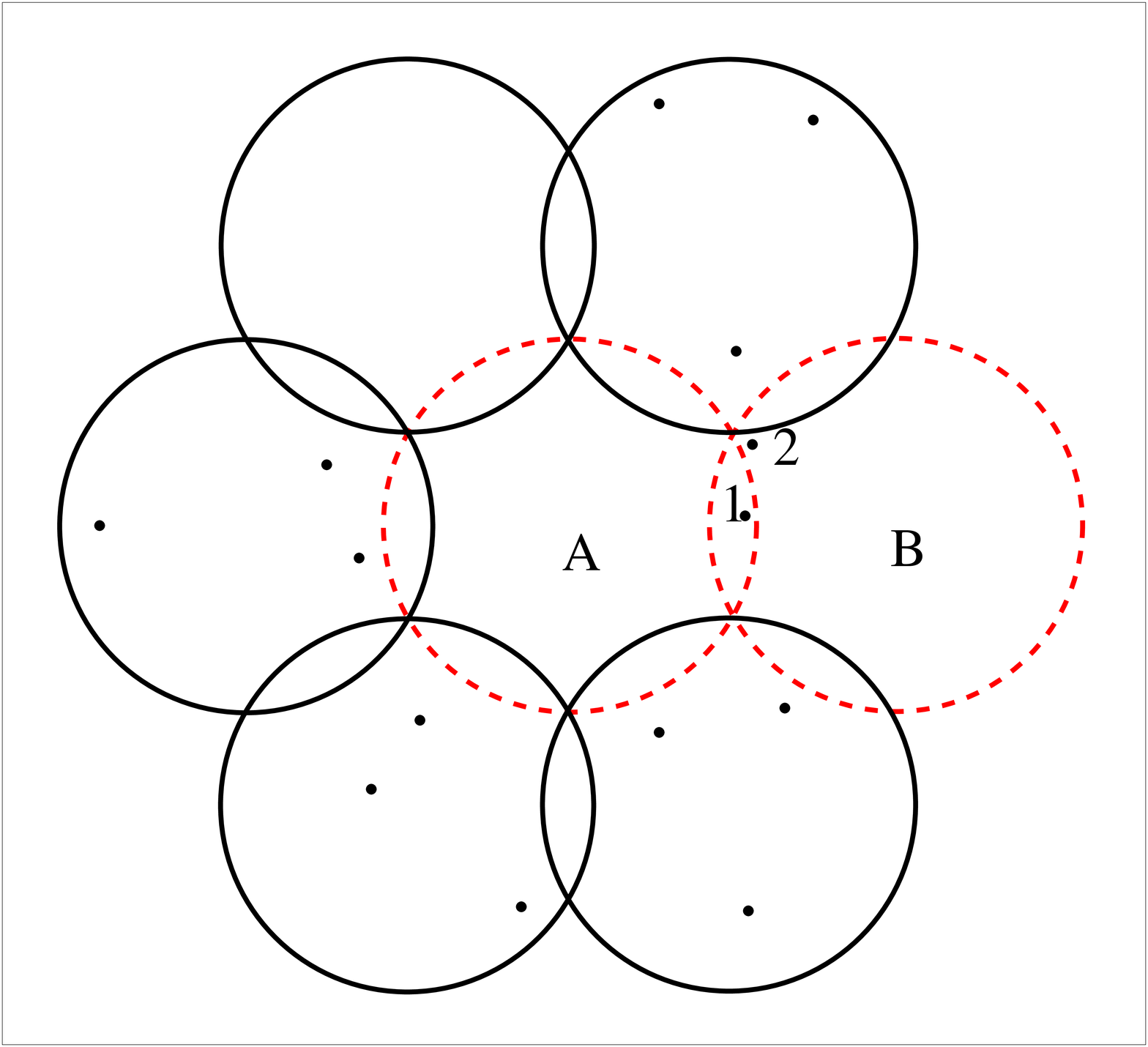}
\caption{Sketch showing an array of 7 patrol discs which illustrates the advantages 
of the draining algorithm presented in this work. Note that an algorithm based on proximity
would assign target 1, which lies in a common area shared by positioner 
A and B, to positioner B, consequently preventing operation
of positioner A. This could also happen with a random assignment. The draining 
algorithm, however, would assign target 1 to positioner A and target 2 to positioner B, thus
causing both targets to be observed in the current tile (consequently 
maximizing the number of assignments).}
\label{fig:cartoon}
\end{figure}

The motivation for optimizing the fiber assignment process derives 
from the fact that a single tile is normally not enough 
to observe all targets at a given pointing, or even a fraction 
of them. At this point and for the sake of understanding how  
the fiber assignment algorithm works, it is convenient to consider tiles as
the different exposures that are needed to observe a given fraction of all targets 
at a given pointing. Tiles should be seen in this context, therefore, as iterations in the fiber assignment process. In practice, 
the number of tiles needed at a given pointing is basically set by $\eta$ and the required completeness. 
As we will see below, only in cases where a very small $\eta$ ($\sim 0.5$) and a relatively low 
completeness ($\sim 80 \%$) are required, will a single iteration (tile) be sufficient. 
These are of course very rare cases. The main goal here is to maximize the fraction of 
all targets observed after a given number of tiles.\\

Several strategies can be developed in order to assign targets to positioners. The simplest approach is to select randomly for each positioner 
any of the targets lying within the corresponding patrol disc. We could also impose a proximity criterion 
for the assignment, as that conceived for LAMOST. Instead of that,
our motivation is to achieve an optimal solution, where optimal here means that    
the maximum number of targets is observed in the first tiles. The idea is to concentrate as many targets as 
possible in the first tile, and then repeat the process for the
second and subsequent tiles until the required completeness is reached. 
We want to move, whenever possible, targets towards
the first tiles, and this is possible because some
of the targets will fall into the common areas of the patrol discs
(where they can be observed by two or, in a few cases, three positioners). In essence, 
the following steps must be implemented:

\begin{itemize}

\item For each target, identify the positioners which can access it.

\item For each positioner, create a list of targets to be observed. In 
case a target can be observed by more than one positioner, it must be
assigned to the positioner with the shortest list. The first target of this list would be
observed in the first tile (iteration), the second target in the second tile and so 
forth.

\item Optimize the assignments. The above configuration can be optimized
by moving targets between different lists (objects can be moved between 
lists if they belong to more than one list, i.e. they are observable by more 
than one positioner). The idea is to shorten the long lists in favor of the shorter ones, so that
all lists tend to the same length.
In practice, we would search list by list for targets that could be moved 
to shorter lists. This process must be iterated until the optimal 
configuration is achieved.

\end{itemize}

To illustrate typical situations where the draining 
algorithm is clearly advantageous we show in Fig.~\ref{fig:cartoon} a simple sketch of an array of 7 positioners, represented by their 
corresponding patrol discs. Note that in a real focal plane each patrol disc overlaps 
with 6 other patrol discs (obviously excluding those in the border). In this example most 
of the targets (dots) sit inside exclusive areas, i.e. regions that are only accessible 
by a single positioner. Target 1, however, lies in the common region shared by positioner A and positioner B. 
An algorithm based on proximity would assign target 1 to positioner B, consequently preventing operation
of positioner A. This could also happen with a random assignment. The draining 
algorithm proposed here, however, would assign target 1 to positioner A and target 2 to positioner B, thus
forcing both targets to be observed in the current tile and consequently 
increasing the number of assignments. Our tests indicate that, over a large number of positioners,
situations like this happen with relevant frequency so the draining algorithm achieves a sensibly better
performance than other approaches, as we will show below. Following 
this idea, it is rather straightforward to identify the conflicting cases and formalize this method into an algorithm so that 
the optimal sequence of tiles is found. \\

 \begin{figure}
\plotone{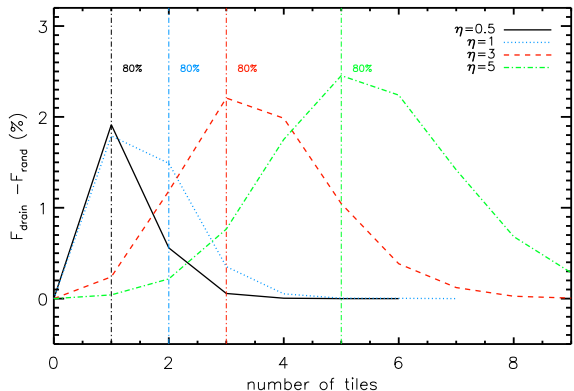}
\caption{Difference in the cumulative fraction of targets assigned with the draining algorithm
        described in Section~\ref{sec:draining} ($F_{\rm drain}$) and with a random approach ($F_{\rm rand}$) shown as a
        function of the number of tiles for different target-to-positioner ratios, $\eta$. Vertical
        lines represent the number of tiles needed to reach the
        $80\%$ completeness level for each $\eta$. Simulations have been performed with 
        100 catalogs where targets are randomly distributed across the focal plane.}
\label{fig:draining}
\end{figure}

\begin{table*}
      \caption{Detailed comparison between the average cumulative fractions of targets assigned with the draining algorithm
        and with a random approach as a function of the number of tiles for different target-to-positioner ratios, $\eta$. For each $\eta$, 
        the tile at which the $80\%$ completeness level is highlighted in bold fonts. Simulations have been performed with 
        100 catalogs where targets are randomly distributed across the focal plane. Standard deviations of the
        cumulative fractions are shown in brackets. We also show results obtained with our draining algorithm 
        combined with rotation of the focal plane, i.e. draining + ROT.}
         \label{tab:random}
         \centering
         \begin{tabular}{c c c c c c}
            \hline
\multicolumn{6}{c}{} \\           

                   Number of tiles & Method  & $\eta=0.5$   & $\eta=1$  & $\eta=3$      &  $\eta=5$   \\

            \hline

            $1$ & Draining   & ${\bf{0.837~(0.027)}}$   & $0.687~(0.013)$       & $0.336~(0.002) $  &  $0.209~(0.001)$  \\
                & Random      & ${\bf{0.817~(0.025)}}$   & $0.669~(0.012)$       & $0.334~(0.002)$  &  $0.208~(0.001)$  \\
                & Draining + ROT   & ${\bf{0.866~(0.027)}}$   & 0.707 (0.011)    & 0.339 (0.001)  & 0.211 (0.000)  \\

            \hline

            2   & Draining   & 0.983 (0.031)        & {\bf{0.937 (0.017)}}      & 0.626 (0.004)  & 0.411 (0.001)   \\
                & Random      & 0.977 (0.030)       & {\bf{0.922 (0.017)}}      & 0.614 (0.004)  & 0.409 (0.001)  \\
                & Draining + ROT   &  0.998 (0.031)   & {\bf{0.966 (0.016)}}   & 0.644 (0.003)   &  0.417 (0.001) \\

          \hline

             3  & Draining   & 0.998 (0.032)   & 0.989 (0.019)  &  {\bf{0.832 (0.007)}}   &  0.599 (0.002)  \\
                & Random      & 0.998 (0.031)   & 0.986 (0.019)  &  {\bf{0.809 (0.007)}}   &  0.592 (0.002) \\
                & Draining + ROT   & 1 (0.031)  & 0.999 (0.017)    & {\bf{0.866 (0.006)}}   & 0.615 (0.001)\\

          \hline
 
             4  & Draining    & 0.999 (0.032)    & 0.998 (0.019)  &  0.940 (0.009)   & 0.762 (0.003)  \\
                & Random       & 0.999 (0.031)    & 0.998 (0.019)  &  0.920 (0.008)   & 0.744 (0.003)  \\
                & Draining + ROT   &    & 1 (0.017)   & 0.971 (0.008)   & 0.788 (0.003)\\

          \hline
  
             5  & Draining    & 0.999 (0.032)   & 0.999 (0.019)  & 0.982 (0.009)   &  {\bf{0.881 (0.005)}}  \\
                & Random       & 0.999 (0.031)   & 0.999 (0.019)  & 0.972 (0.009)  &  {\bf{0.856 (0.006)}}  \\
                & Draining + ROT   &    &    & 0.998 (0.009)   & {\bf{ 0.913 (0.004)}}\\

          \hline

             6  & Draining    & 1 (0.032)    & 0.999 (0.019)  & 0.995 (0.010)   &  0.950 (0.006)  \\
                & Random       & 1 (0.031)   & 0.999 (0.019)  & 0.991 (0.009)  &  0.928 (0.006) \\
                & Draining + ROT   &    &    & 1 (0.009)  &  0.978 (0.006)\\

          \hline
  
             7  & Draining    &     & 1 (0.019)  & 0.999 (0.010)   & 0.981 (0.007)    \\
                & Random       &     & 1 (0.019)  & 0.997 (0.009)  & 0.967 (0.006)   \\
                & Draining + ROT   &    &    &    & 0.997 (0.006)\\

          \hline

             8  & Draining    &     &   & 0.999 (0.019)   & 0.994 (0.007)   \\
                & Random       &     &   & 0.999 (0.009)   & 0.987 (0.006)   \\
                & Draining + ROT   &    &    &    & 0.999 (0.006) \\

          \hline
   
             9  & Draining    &     &   & 0.999 (0.010)   &  0.998 (0.007)  \\
                & Random       &     &   & 0.999 (0.009)  &  0.995 (0.006)  \\
                & Draining + ROT   &    &    &    & 1 (0.006)\\
  
          \hline
  
             10 & Draining     &     &   & 0.999 (0.010)   & 0.999 (0.007)   \\
                & Random       &     &   & 0.999 (0.009)  & 0.998 (0.006)  \\
                & Draining + ROT   &    &    &    & \\
 
            \hline
         \end{tabular}
   \end{table*}

\begin{figure}
\plotone{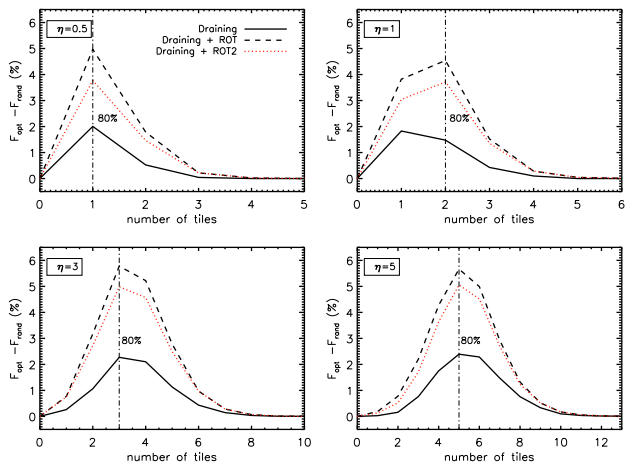}
\caption{Cumulative percentage of targets gained with respect to a random 
assignment by using the draining algorithm described in Section~\ref{sec:draining} alone (solid line) and 
the same algorithm allowing rotation of the focal plane as discussed in Section~\ref{sec:additional} (including ROT and ROT2), as a function 
of the number of tiles for different target-to-positioner ratios, $\eta$. Vertical
lines represent the number of tiles needed to reach the
$80\%$ completeness level for each $\eta$. Simulations have been performed with 
100 catalogs of randomly distributed targets.}
\label{fig:random}
\end{figure}

In order to assess the performance of the draining algorithm, we have generated 
a set of 100 catalogs of randomly distributed targets for systems with four different target-to-positioner ratios, 
namely $\eta = 0.5,1,3,5$. Note that, as shown in Table~\ref{tab:BB_SIDE}, 
the values of $\eta = 1$ and $\eta = 5$ correspond to the two examples considered in this paper, i.e. SIDE and BigBOSS, respectively. 
We have implemented both the draining algorithm and a simple approach where targets are 
randomly assigned to positioners. In Table~\ref{tab:random}, we show the average total fraction of 
all targets assigned to positioners after each of the first 10 tiles (cumulative fractions) using both these methods, 
for the target-to-positioner ratios considered. The tile at which $80\%$ of the total number of targets 
is successfully assigned to positioners has been highlighted for each $\eta$. The number of tiles 
needed to reach this completeness (that we have selected arbitrarily) obviously increases 
with the target-to-positioner ratio, ranging from only 1-2 tiles for $\eta \leq 1$ to up to 5 tiles for $\eta=5$. At this
point it is necessary to remind that the scope of this optimization is not to reduce drastically the average number of tiles
needed to achieve a certain completeness, but to increase the fraction of targets observed in the same number of tiles. As we can see 
in Table~\ref{tab:random}, the draining algorithm presented here increases the assignments in approximately $2\%$ as compare to a greedy 
procedure at the $80 \%$ completeness level. \\

\begin{figure*}
\begin{center}
\begin{tabular}{c}
\includegraphics[scale=0.7]{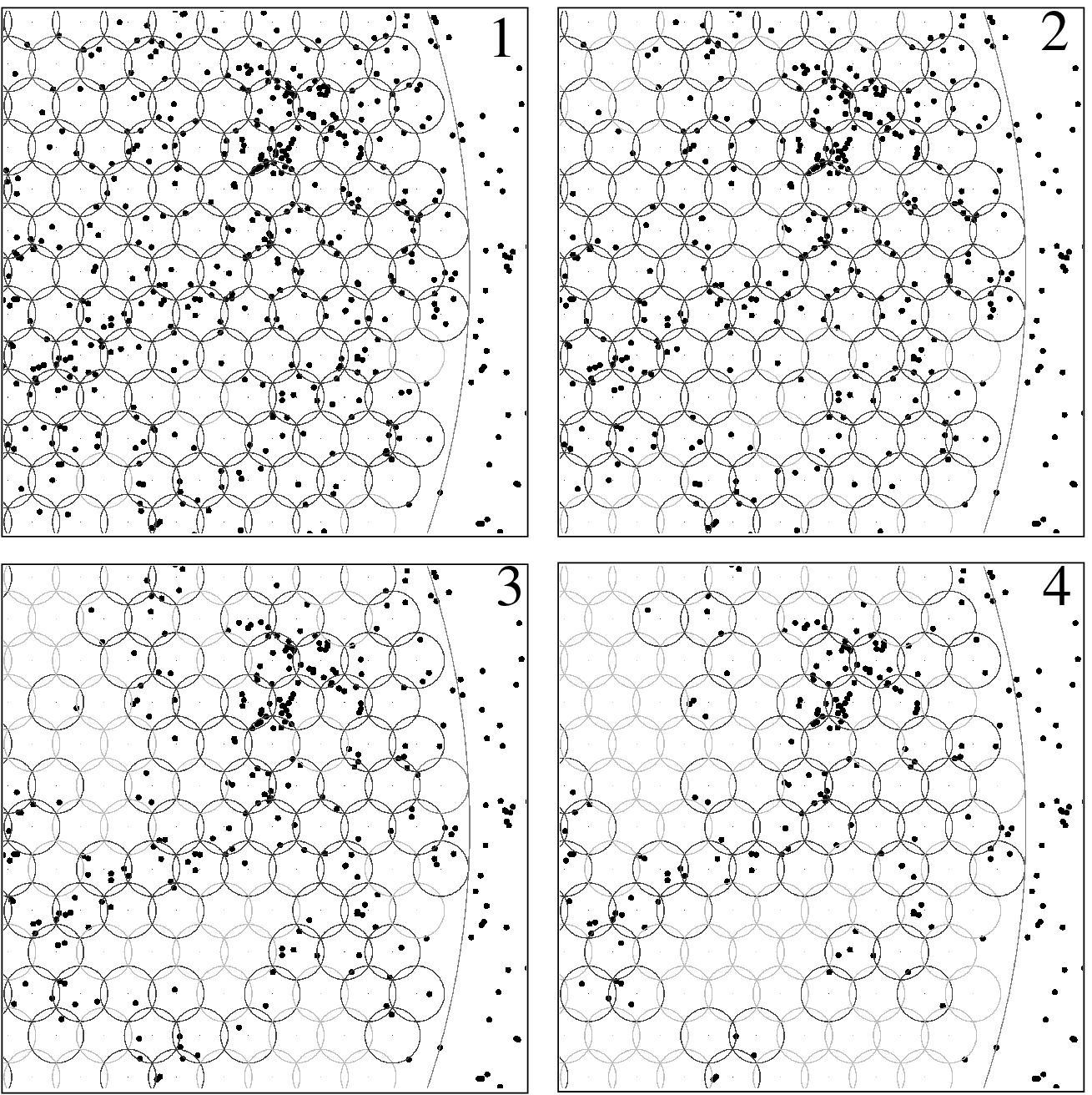}
\end{tabular}
\end{center}
\caption{Portion of the focal plane of an instrument, represented by 
an array of patrol discs, in a sequence of the first tiles. The over-plotted dots 
symbolize the projected positions of targets belonging to a mock galaxy catalog with a target-to-positioner ratio $\eta \sim 3$.
The first panel in this figure (upper-left panel) illustrates the initial situation where the entire sample of 
targets is to be observed. Patrol discs with at least one target inside have been highlighted. The draining 
algorithm maximizes the number of targets observed in each tile. In this example, $\sim 75\%$ of targets 
have been selected after 3 tiles.}
\label{fig:mocks}
\end{figure*}

Fig.~\ref{fig:draining} also illustrates the performance of the draining algorithm as 
compared to a random assignment. Here, the difference 
in the cumulative percentage of targets assigned with our optimized algorithm ($F_{drain}$)
and with a random approach ($F_{rand}$) is shown as a function of the
number of tiles. Fig.~\ref{fig:draining} shows that the gain provided by this optimized method
increases slightly with $\eta$, ranging from $\sim 1.7\%$ for 
$\eta=0.5$ to $\sim 2.5\%$ for $\eta=5$. Note that, due to the nature of this optimization, 
the improvement occurs in the first few tiles. In the following sections we will show how an improvement like this 
can have a strong effect in the design of a survey.\\

An important issue related to the fiber assignment process in this 
type of fiber-fed spectrographs is the so-called fiber collision problem. This concept was  
used to describe the fact that in the SDSS \citep{York2000} fibers could not 
be placed closer than $55''$ arcsec. Spectrographs that feature a fiber positioning 
robot as that described in Section~\ref{sec:robot} present 
a similar problem, as due to the physical size of the positioners, frequently 
two or more targets cannot be observed simultaneously (i.e. in the same tile). The results shown 
in this work were obtained assuming that positioners have no physical size, hence
neglecting the possibility of fiber collisions. The typical number of collisions as a function of the target-to-positioner
ratio, $\eta$, will depend on the geometry of the system, i.e. on the number and the size of the 
positioners. The problem, however, can be solved naturally within the framework
of the draining algorithm. Fiber collisions will occur exclusively in common areas and this method
is specifically designed to deal with objects in these regions conveniently, and to ensure that
the maximum number of targets is observed in the first tiles. We have checked that the 
draining algorithm alone can solve naturally almost all collisions, so the results presented in 
Fig.~\ref{fig:draining} and Table~\ref{tab:random} are not affected by our assumption.  
Note that the number of these events in catalogs of randomly distributed targets is absolutely negligible. 
Even in real catalogs, where fiber collisions are more frequent, the draining algorithm   
provides an optimal solution, as we will show in Section~\ref{sec:real}.

\section{Additional optimizations}
\label{sec:additional}

So far we have presented an optimized method for fiber positioning, the draining algorithm, that 
can be easily implemented in any future fiber-fed spectrograph. 
In this section we explore additional ways to optimize the number of targets observed that require
some pre-designed capabilities from our instrument. In particular, we will 
demonstrate that it is possible to achieve a much better performance by 
simply allowing the focal plane of the instrument to rotate.\\

In Section~\ref{sec:nom} we defined the position angle, PA, describing the rotation 
of our focal plane. Note that an array of positioners as that described in Section~\ref{sec:robot} is geometrically symmetric 
under PA rotations of $60^{o}$ (see an example in Fig.~\ref{fig:focalplane}). 
Let us first assume that our instrument is capable of rotating the focal 
plane up to this angle and consequently all possible configurations are 
accessible. Rotating the focal plane implies that the number of targets
sitting inside each patrol disc changes. More importantly, it changes 
the number of positioners with at least one target within the patrol disc. It seems convenient,
therefore, to find the configuration that optimizes the efficiency 
of the assignment process. In this section, we combine this position angle optimization (ROT optimization)
with the draining algorithm (draining + ROT).\\

 \begin{table*}
      \caption{The same as in Table 2 but using mock catalogs extracted from the Bolshoi Simulation. For each target-to-positioner 
ratio, $\eta$, we used a set of 9 catalogs.}
         \label{tab:real}
         \centering
         \begin{tabular}{c c c c c c}
            \hline
\multicolumn{6}{c}{} \\           

                   Number of tiles  & Method  & $\eta=0.5$   & $\eta=1$  & $\eta=3$      &  $\eta=5$   \\

            \hline

            $1$     & Draining                        & 0.737       & $0.599$       & $0.321$  &  $0.207$  \\
                    & Random                           & 0.718       & $0.585$       & $0.319$  &  $0.206$  \\
                    & Draining + ROT             & 0.766       & $0.622$       & $0.327$  &  $0.208$  \\

            \hline

            2   & Draining               & {\bf{0.934}}    & {\bf{0.855}}   & 0.576  & 0.398   \\
                & Random                 & {\bf{0.923}}    & {\bf{0.840}}   & 0.566  & 0.394   \\
                & Draining + ROT   & {\bf{0.969}}    & {\bf{0.897}}   & 0.599  & 0.406   \\
 
          \hline

             3  & Draining              & 0.983   & 0.950  & 0.755   &  0.565  \\
                & Random                 & 0.978   & 0.940  & 0.738   &  0.556  \\
                & Draining + ROT   & 0.999   & 0.980  & 0.791   &  0.584  \\
 
          \hline
 
             4  & Draining               & 0.996    & 0.984  & {\bf{0.863}}   & 0.701  \\
                & Random                  & 0.995    & 0.979  & {\bf{ 0.847}}  & 0.687  \\
                & Draining + ROT    & 1        & 0.998  & {\bf{ 0.905}}   & 0.730  \\

          \hline
  
             5  & Draining               & 0.9999    & 0.994  & 0.926   &  {\bf{0.803}}  \\
                & Random                  & 0.9999    & 0.993  & 0.913   &  0.786  \\
                & Draining + ROT   &           & 1      & 0.962   &  {\bf{0.838}}  \\

          \hline

             6  & Draining               & 1    & 0.998       & 0.960   &  0.874  \\
                & Random                  & 1    & 0.997       & 0.951   &  {\bf{0.856}}  \\
                & Draining + ROT    &      &             & 0.986   &  0.909  \\

          \hline
  
             7  & Draining               &     & 1           & 0.978   & 0.921   \\
                & Random                  &     & 0.999       & 0.972   & 0.905   \\
                & Draining + ROT    &     &             & 0.995   & 0.952   \\

          \hline

             8  & Draining               &     & 0.999   & 0.988       & 0.950   \\
                & Random                  &     & 0.999   & 0.984       & 0.938   \\
                & Draining + ROT    &     &         & 0.998       & 0.976   \\

          \hline
   
             9  & Draining               &     & 1         & 0.993      &  0.968  \\
                & Random                  &     & 1         & 0.991      &  0.959  \\
                & Draining + ROT    &     &           & 0.999      &  0.989  \\
 
          \hline
  
             10 & Draining                &     &          & 0.996      & 0.980   \\
                & Random                   &     &          & 0.995      & 0.972   \\
                & Draining + ROT     &     &          & 0.999      & 0.995   \\

            \hline
         \end{tabular}
   \end{table*}

The draining algorithm is a simple method which optimizes the fiber assignment process by
ensuring that the maximum number of targets is assigned in the first tiles. Such a method provides an 
optimized solution for our observation plan that normally includes several tiles
, depending on $\eta$ and the completeness requested. The easiest way to insert the ROT optimization
into this scheme is to simply allow for rotation at the beginning of the process and find the 
configuration that maximizes the number of positioners with at least one target in the
patrol disc. However, it is obviously much more efficient to permit rotation not only at the beginning 
of the process but also between tiles (iterations of the process). This ensures that the maximum number of positioners is in 
use in each tile and requires a trivial modification of the above scheme. Now, the draining algorithm is 
applied ``from scratch'' in each iteration of the process on the targets left unobserved from the previous iteration 
and after the optimal PA is found. Obviously, only the first tile from 
the solution provided by our optimized algorithm will eventually be performed for each step.   
This does not jeopardize the proper optimization of the process by any means, as the algorithm is 
specifically designed so that the maximum number of assignments is achieved in the first tile.\\

We have assessed the relevance of the ROT optimization in the same 100 catalogs of 
randomly distributed targets that we used in Section~\ref{sec:draining}. In order to isolate 
the effect of rotation on the efficiency of the process, we need to conveniently
address what we call the ``border scan effect''. Usually the target field 
is a circle on the sky encompassing the most external patrol discs of the array of positioners.
As we rotate the focal plane, some of the targets lying between the outermost perimeter of the
patrolled area, which is not a circle, and the border of the target field can become accessible to 
some positioners. It is important not to overestimate the improvement by 
including targets which would be out of reach of the patrolled area if no rotation 
were applied. In order to avoid this ``contamination'', 
we have restricted ourselves to an area of the focal plane where the total number 
of targets remains constant under any rotation.\\

In Table~\ref{tab:random} we presented the average cumulative fractions of targets assigned with the draining algorithm 
and with a random approach in the first 10 tiles for different values of $\eta$. We also
show for comparison the same fractions obtained using the combined draining + ROT algorithm described in this section. 
Interestingly, allowing for rotation reduces drastically the total number of tiles needed, namely only 4 instead of 7 for 
 $\eta=1$, as an example. This is not critical as far as a real observation is concerned, unless a $100\%$
completeness were required. More importantly, the increase in the efficiency 
of the assignment in the relevant tiles provided by the PA optimization is very significant as compared to 
the case were rotation is not permitted: $3-4 \%$ even if we were to use the draining algorithm and 
$5-6\%$ if we adopted a greedy aproach. This improvement is shown in a clear way 
in Fig.~\ref{fig:random}, which is similar to Fig.~\ref{fig:draining} but including the cumulative percentage of targets gained with respect 
to a random approach when rotation of the focal plane is allowed.\\

There are instruments where only slight rotations of the focal plane
are permitted. As an example, the BigBOSS spectrograph is expected to
be capable of rotating its focal plane by no more than $\sim 2^{o}$.
We have studied the effect that a limited ROT optimization would have
in the efficiency of the fiber assignment process by allowing our
simulated focal plane to rotate a maximum of $\pm 2^{o}$ (ROT2
optimization). In Fig.~\ref{fig:random} we show in a dotted line the result of
applying this test on our set of random catalogs. Interestingly, even
a small rotation like this produces a remarkable optimization: 
$3.5-5\%$ at a completeness level of $80\%$. Also, an alternative 
to the ROT optimization for instruments where rotation of the 
focal plane is not permitted might be found in tweaking the telescope 
pointing position around the center of the target field. Again, this should lead to configurations
where targets are more conveniently spread over the array of positioners, consequently increasing the efficiency 
of the assignment process. It is therefore reasonable to expect an optimization like this to produce
similar results as those found with the ROT optimization. \\

\section{Application to mock galaxy catalogs}
\label{sec:real}

In previous sections we discussed the performance of our optimized algorithm for 
fiber positioning (including rotation or, alternatively, small shifts of the focal plane of our instrument) when 
implemented in catalogs of randomly distributed targets. However, it is well known that 
galaxies in the Universe are far from being randomly distributed. Galaxies are actually clustered, 
forming filaments in the 3-dimensional space that are separated from each other by under-dense regions. 
In a similar manner would targets be distributed in the 2-dimensional projection of this space onto
the focal plane of our instrument. Consequently, in a realistic situation, targets 
would accumulate in certain regions of our focal plane, leaving others relatively underpopulated. We have simulated
the performance of our optimizations in real-life situations by making use of galaxy mock catalogs extracted from the 
Bolshoi simulation \citep{Klypin2010}. Important for this work, the clustering properties 
of mock galaxies match those of real galaxies with good accuracy. The reason for using mock catalogs 
instead of real catalogs is that they provide us with more flexibility when it comes to selecting 
different samples with different number densities. We have performed a 2-D projection of a 
simulation box of 250 Mpc/h on a side situated at redshift 1 onto our focal
plane and used the circular velocity of haloes as an empirical threshold for selecting different densities. In order 
to allow for a fair comparison with the results presented in the previous sections, we have selected 
catalogs with target-to-positioner ratios of $\sim$ 0.5, 1, 3 and 5. Simulations were performed with 
9 different realizations for each $\eta$, obtained by rotating the box conveniently.\\

In Fig.~\ref{fig:mocks}, we show a portion of the focal plane of an instrument, represented by 
an array of patrol discs, in a sequence of the first tiles. The over-plotted dots 
symbolize the projected positions of targets belonging to a mock galaxy catalog with $\eta \sim 3$.
The first panel in this figure illustrates the initial situation where the entire sample of 
targets is to be observed. Note that, as explained above, targets on the focal 
plane of the spectrograph, rather than being randomly distributed, accumulate in filaments. The rest 
of the panels show the distribution of targets assigned to each tile by using the
optimized algorithm described in Section~\ref{sec:draining}. In this example, $\sim 75\%$ of targets 
have been selected after 3 tiles.\\

\begin{figure}
\plotone{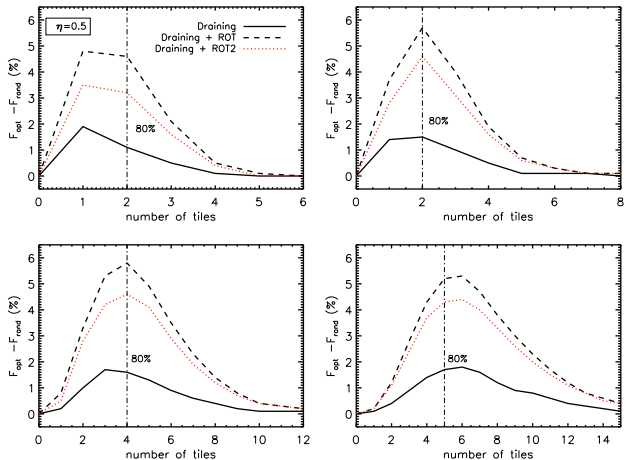}
\caption{The same as in Fig.~\ref{fig:random} but using galaxy mock catalogs extracted from the Bolshoi simulation.
         For each target-to-positioner ratio, $\eta$, 9 mock catalogs were used.}
\label{fig:real}
\end{figure}

The performance of our optimizations with mock galaxy catalogs is presented in the same 
format as in previous sections in Table~\ref{tab:real} and Fig.~\ref{fig:real}. A direct consequence of the 
presence of clustering in our catalogs is that the fraction of targets assigned to each tile
decreases significantly, as a comparison between Table~\ref{tab:random} and Table~\ref{tab:real} demonstrates. Trivially, the probability 
that a single positioner has to deal with several targets is now higher and, consequently, 
it becomes harder to move targets towards the first tiles. As an example, with $\eta \sim 1$ and two tiles
we could observe almost $94 \%$ of all targets in a random catalog, even whithout 
allowing for rotation of the focal plane. In a real catalog we could only assign $85 \%$ of targets in 
the same number of tiles. Similarly, in a real catalog with $\eta \sim 5$ we would need 5 tiles to 
barely reach a completeness of $80 \%$, at least $8\%$ below our expectations 
from random catalogs. Again, we refer the reader to Table~\ref{tab:real} for the exact fractions 
of targets assigned with a random approach, with the draining algorithm alone and 
with the draining algorithm complemented with rotation of the focal plane. In order 
to analyze the performance of our optimizations as compared to a random approach we point 
the reader to Fig.~\ref{fig:real}. This figure shows that the gain provided by our optimizations as compare to a greedy approach 
when implemented in a real-life situation is consistent with what we could infer from random catalogs. A closer 
inspection, however, reveals that the gain provided by the draining algorithm alone is slightly smaller now, falling below $2\%$, whereas
the gain achieved by combining this algorithm with a ROT optimization remains in the range of $5-6\%$. If only 
slight rotations were allowed (ROT2 optimization) we could still improve the efficiency of the process in $3.5-4.5\%$.\\

Note that, as mentioned previously, the results shown in this work were obtained ignoring fiber collisions, as
the effect of these depends on the geometry of the fiber positioning robot itself, and, therefore, cannot
be extrapolated to any fiber-fed spectrograph of this kind. The typical number 
of these events obviously increases in mock galaxy catalogs (and hence in real catalogs) due to the 
fact that objects are more clustered (the fraction of collisions is almost negligible in random catalogs).
Even in real catalogs the fraction of objects in conflict remains small for SIDE: $\lesssim 1\%$ even for $\eta=5$. 
However, an important advantage of the draining algorithm is that collisions can be solved optimally, 
and additional methods are not needed. The results presented in Table~\ref{tab:real} and Fig.~\ref{fig:real} would basically not change 
if collisions were taken into account. Only very slight variations are expected in the very last tiles, which are 
not relevant in real-life observations. \\   

The results presented in this section confirm that, despite the physical restrictions of 
this state-of-the-art fiber positioning robots, it is possible to optimize a survey strategy in 
a remarkable way just by assigning targets to tiles and positioners conveniently. In addition, these results 
represent a strong support for allowing the focal plane of future instruments to rotate. We have shown how
even a slight rotation produces a remarkable optimization. In the next section we discuss on the implications of our results.

\section{Discussion and Conclusions}
\label{sec:discussion}

The results on the optimization of the fiber positioning process that we present in this work are 
valid for any focal plane consisting of an array of positioners as that described in Section~\ref{sec:robot}. To first order, therefore,
our results are not dependent on the size of the tile or the number of fibers per tile, as long 
as positioners are arrayed in a hexagonal pattern as that shown in Fig.~\ref{fig:focalplane}. This is a 
standard concept in state-of-the-art fiber-fed spectrographs, such as some of the ones proposed for the 
GTC or others like BigBOSS, LAMOST, etcetera. Let us illustrate the relevance of our 
results with a simple example based on the future survey BigBOSS. BigBOSS is planned to 
map an area of approximately $14,000$ sq. deg. The expected target density reportedly hovers around
$3,500~targets/deg^{2}$ with a required completeness of $\sim 80\%$. This would yield a 
huge survey, of a few tens of millions of objects. According to our estimations, 
by using the draining algorithm presented here instead of a random assignment, we could save for observation $\sim 2\%$
of all targets, that is, several hundred thousand objects (expectedly $> 700,000$). Note that
a sample like this would by itself triple the size of the 2dF Galaxy Redshift Survey \citep{Colless2001} 
and would be comparable to the SDSS. If we could take advantage 
of the fact that the BigBOSS instrument allows for slight rotations of the focal plane, the above gain 
would increase in at least a factor $2$, which represents a remarkable improvement. The increase in the 
efficiency of the fiber positioning process that these optimizations provide could also
reduce the costs of the survey, in terms of observing time. In fact, by simply implementing 
the draining algorithm and according to our estimations, we could collect the same
number of objects in a remarkably smaller area: between $\sim 350~deg^{2}$ and $\sim~700~deg^{2}$ smaller
depending on the feasibility of rotation. Assuming that we need an average of 5 tiles per pointing, that 
would mean saving between $\sim 50$ and $\sim 100$ tiles.\\

\begin{figure}
\plotone{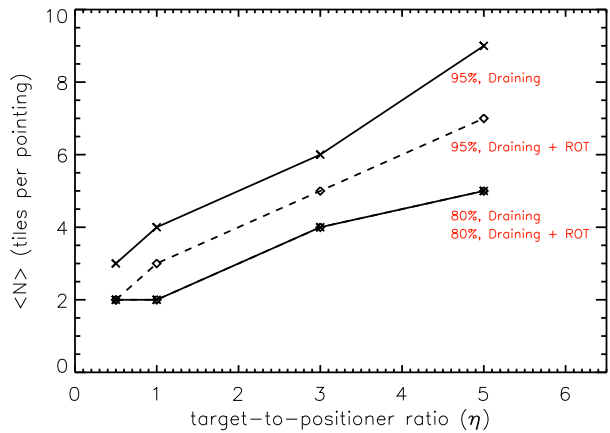}
\caption{Average number of tiles per pointing, $<N>$, as a function
of the target-to-positioner ratio, $\eta$, for 2 different completeness levels: $80 \%$ and $95 \%$}
\label{fig:discuss}
\end{figure}

The intention of this work is not only to present some remarkable fiber positioning optimizations but also to provide 
some basic guidelines for an efficient instrument requirement definition for fiber-fed multi-object spectrographs. Fig.~\ref{fig:discuss} is 
intended to summarize some of our main results in a way that may be useful for future surveys. In this 
figure we show the average number of tiles per pointing, $<N>$, as a function
of the target-to-positioner ratio, $\eta$, for 2 different completeness levels: $80 \%$ and $95 \%$. These
values define a reasonable range for the completeness required in a real survey. Note that 
$\eta$ and the completeness requested determine completely (at least to first order) the average
number of tiles per pointing and, consequently, the (approximate) tile density of our survey. The 
parameter $\eta$ is the ratio of the target density to the fiber density. The target density is a requirement 
that strongly depends on the science case. The fiber density is restricted by current technology as there is
a limit in the size of the positioner. Ideally, we want to increase our fiber density as much as possible,
consequently decreasing $\eta$, so that fewer tiles are needed to complete our survey. This obviously 
reduces the observation time and consequently the cost of the survey.\vspace{0.25cm}

Finally, we summarize the main results of this work as follows:\vspace{-0.15cm}
\begin{itemize}
\item We have presented an optimized algorithm for assigning fibers to targets in 
next-generation fiber-fed multi-object spectrographs: the draining algorithm. Our method 
is very easy to implement and ensures that the maximum number of targets in
a given target field is observed in the first few tiles. Using both catalogs of randomly 
distributed targets and mock galaxy catalogs drawn from cosmological simulations 
we have estimated that the 
gain provided by the draining algorithm as compared to a random assignment
can be as much as $2\%$ for the first tiles.

\item The fiber collision problem can be solved easily and in an optimal 
way within the framework of the draining algorithm.

\item We also discuss additional optimizations of the fiber positioning process. In particular,
we have shown that allowing for rotation of the focal plane can improve the efficiency 
of the process in $\sim 3.5 - 4.5 \%$ even if only small adjustments are permitted (up 
to $2^{o}$, ROT2 optimization). For instruments that allow large rotations of the focal plane 
(ROT optimization) the expected gain increases to $\sim 5 - 6 \%$. These results, therefore, strongly support 
focal plane rotation in future spectrographs, as far as the efficiency of the
fiber positioning process is concerned

\item An alternative to the ROT optimization for instruments where rotation of the 
focal plane is not permitted might be found in tweaking the telescope 
pointing position around the center of the target field. We expect an optimization like this to produce
similar results as those found with the ROT optimization.

\item As an example of a future large-scale galaxy survey, we have discussed the implications
of our optimizations if applied to BigBOSS ($\sim 14,000~deg^{2}$, $\sim 40 M$ objects). We have estimated that by using the draining 
algorithm instead of a random assignment we could save for observation several hundred 
thousand objects (expectedly $> 700,000$, a sample comparable to the SDSS). Alternatively, we could collect the total
number of objects expected for the entire survey in a remarkably smaller area: $\sim 350~deg^{2}$ smaller 
(equivalent to $\sim 50$ tiles). This figures would typically double if we could
perform focal plane rotations of at least $2^{o}$.  

\item We provide the average tile density as a function of the target-to-positioner 
ratio, $\eta$, and completeness, which are the fundamental quantities that defines a survey or the instrument requirements. These 
results are applicable to most next-generation fiber-fed multi-object spectrographs. 
\end{itemize}

\section*{Acknowledgments}

We thank the support of the Spanish MICINN's Consolider-Ingenio 2010  
Programme under grant MultiDarkCSD2009-00064. We also acknowledge  
support from the MICINN's grant AYA2010-21231 and the CSIC 
IAA-LBNL i-LINK0040 grant. We thank J. Betancort-Rijo for 
his contribution in the early phase of this work.
We are grateful to A. Klypin for providing Bolshoi catalogs.
We thank M. Blanton and D. Schlegel for stimulating discussions
and providing comments on the text and results.

\bibliography{./paper}

\begin{thebibliography}{13}
\expandafter\ifx\csname natexlab\endcsname\relax\def\natexlab#1{#1}\fi

\bibitem[{Azzaro} et~al.(2010){Azzaro}, {Becerril}, {Vilar} et~al.]{Azzaro2010}
{Azzaro} M., {Becerril} S., {Vilar} C., et~al., 2010, in { Society of
  Photo-Optical Instrumentation Engineers (SPIE) Conference Series\/}, vol.
  7735 of { Presented at the Society of Photo-Optical Instrumentation Engineers
  (SPIE) Conference\/}

\bibitem[{Bassett} et~al.(2005){Bassett}, {Nichol} \&
  {Eisenstein}]{Bassett2005}
{Bassett} B.~A., {Nichol} B., {Eisenstein} D.~J., 2005, Astronomy and
  Geophysics, 46, 5, 050000

\bibitem[{Bell} et~al.(2009){Bell}, {Davis}, {Dey} et~al.]{Bell2009}
{Bell} E., {Davis} M., {Dey} A., et~al., 2009, in { astro2010: The Astronomy
  and Astrophysics Decadal Survey\/}, vol. 2010 of { ArXiv Astrophysics
  e-prints\/},  arXiv:0903.3404

\bibitem[{Blanton} et~al.(2003){Blanton}, {Lin}, {Lupton} et~al.]{Blanton2003}
{Blanton} M.~R., {Lin} H., {Lupton} R.~H., et~al., 2003, \aj, 125, 2276

\bibitem[{Colless} et~al.(2001){Colless}, {Dalton}, {Maddox}
  et~al.]{Colless2001}
{Colless} M., {Dalton} G., {Maddox} S., et~al., 2001, \mnras, 328, 1039

\bibitem[{Huchra} et~al.(1983){Huchra}, {Davis}, {Latham} \&
  {Tonry}]{Huchra1983}
{Huchra} J., {Davis} M., {Latham} D., {Tonry} J., 1983, \apjs, 52, 89

\bibitem[{Kimura} et~al.(2010){Kimura}, {Maihara}, {Iwamuro}
  et~al.]{Kimura2010}
{Kimura} M., {Maihara} T., {Iwamuro} F., et~al., 2010, \pasj, 62, 1135

\bibitem[{Klypin} et~al.(2010){Klypin}, {Trujillo-Gomez} \&
  {Primack}]{Klypin2010}
{Klypin} A., {Trujillo-Gomez} S., {Primack} J., 2010, ArXiv e-prints,
  arXiv:1002.3660

\bibitem[{Peacock} \& {Schneider}(2006)]{Peacock2006}
{Peacock} J., {Schneider} P., 2006, The Messenger, 125, 48

\bibitem[{Prada} et~al.(2008){Prada}, {Azzaro}, {Rabaza}, {S{\'a}nchez} \&
  {Ubierna}]{Prada2008}
{Prada} F., {Azzaro} M., {Rabaza} O., {S{\'a}nchez} J., {Ubierna} M., 2008, in
  { Society of Photo-Optical Instrumentation Engineers (SPIE) Conference
  Series\/}, vol. 7014 of { Presented at the Society of Photo-Optical
  Instrumentation Engineers (SPIE) Conference\/}

\bibitem[{Schlegel} et~al.(2009){Schlegel}, {Bebek}, {Heetderks}
  et~al.]{Schlegel2009}
{Schlegel} D.~J., {Bebek} C., {Heetderks} H., et~al., 2009, ArXiv e-prints,
  arXiv:0904.0468

\bibitem[{Wang} et~al.(2009){Wang}, {Chen}, {Zheng}, {Wu}, {Zhang} \&
  {Zhao}]{Wang2009}
{Wang} X., {Chen} X., {Zheng} Z., {Wu} F., {Zhang} P., {Zhao} Y., 2009, \mnras,
  394, 1775

\bibitem[{York} et~al.(2000){York}, {Adelman}, {Anderson} et~al.]{York2000}
{York} D.~G., {Adelman} J., {Anderson} Jr. J.~E., et~al., 2000, \aj, 120, 1579

\end{thebibliography}

\label{lastpage}

\end{document}